    \documentstyle[epsf]{aipproc}

\def\etal{{\it et al.\ }}

\begin{document}  
    \title{Gazing into the MgF$_2${\thanks {MgF$_2$ is transparent to UV, unlike
    silicate-based glasses}} 
    ball: UV astronomy for the 3rd millenium}  
    \author{Noah Brosch$^{\dagger}$}  
    \address{$^\dagger$Wise Observatory and School of Physics and Astronomy, 
    Raymond and Beverly Sackler Faculty of Exact Sciences, Tel Aviv 
    University, Tel Aviv 69978, Israel} 
    \maketitle  
 
    \begin{abstract} 
 
    I examine possible futures for UV astronomy, in view of the past history 
    and collected knowledge of the UV sky through various missions since the 
    early days of Space Astronomy. The last all-sky survey has been by 
    TD-1 in the early 1970s, resulting in reasonable mapping of discrete sources 
    brighter than ~9th mag (monochromatic, UV). Since TD-1, few advances were made in 
    general knowledge of the UV through survey missions, but dedicated, 
    single-object studies advanced mainly through the IUE spectroscopic
    results. Selected sky 
    areas were surveyed to deeper magnitudes than by TD-1 with various 
    balloon and rocket flights, Shuttle and MIR payloads, etc. The 
    HST offered some UV capability with WFPC 1 and 2, FOC, and presently 
    with STIS, but with small fields of view.  
 
    The future of UV astronomy seems bleak, with only FUSE with any 
    certainity of being flown and with the recent recommendation of the 
    Dressler committee to emphasize NIR and interferometry into the next 
    millenium. The community's hopes  reside therefore with (a) better exploitation of 
    existing data bases, (b) dual-usage missions, (c) piggy-back modern 
    payloads, and (d) low cost alternatives such as long-duration balloon
    flights. These could demonstrate 
    the advantage of a real UV sky survey mission, equivalent in the number 
    of sources to the POSS I \& II and ESO optical surveys.  
 
    \end{abstract}  
 
    \section*{Introduction}  
    We are only a few years away from the beginning of the third millenium. 
    This is a propitious instance to review the status of UV astronomy, and 
    to predict how it is likely to develop in the next few decades. 
    I concentrate here on the spectral band from $\sim$100 nm to 
    $\sim$350 nm; the lower limit is near the border between the 
    extreme UV (EUV) and the regular UV, between imaging optics which work 
    by grazing incidence reflection and normal incidence telescopes. The 
    long wavelength limit is close to the location where ground-based 
    telescopes begin to be efficient.  
 
    In this paper I   express the brightness of an object in 
    ``monochromatic magnitudes'', defined as:  
    \begin{equation}  
    m_\lambda=-2.5 log(F_\lambda)-21.175  
    \end{equation}  
    where F$_{\lambda}$ is the  source flux density at
    $\lambda$ in erg sec$^{-1}$ cm$^{-2}$ \AA$^{-1}$.  
 
    The outline of this paper is as follows: first, the achievements of UV 
    astronomy are reviewed in very broad terms. I concentrate 
    mainly on imaging and photometric results, and  mention briefly 
    a few spectroscopic missions. I review  then  some critical decisions 
    about the future developments of Astronomy, taken by the US astronomical 
    community and adopted by NASA and the US government. These decisions 
    serve as templates for other national agencies and, as will become clear 
    below, do not bode well for the UV domain. I try to understand 
    why those decisions were taken, and  sketch possible ways of 
    reversing the negative trend toward UV in the community at large.

    \section*{UV astronomy: the promise}
  
    The UV range, from $\sim$250 nm down to the limit considered her, is almost
    certainly the region where the sky is the darkest. This has been pointed 
    out by O'Connell (1987) and has since been revised a number of times. 
    Recently, the UV background (UVB) was evaluated from the UIT images 
    (Waller {\it et al.} 1995),  the FAUST data (Sasseen \etal 1995), and  
    the long exposure Voyager UVS spectra (Murthy {\it et al.} 1996). The new evaluations 
    were done at different wavelengths and require correction for Milky Way 
    UV photons scattered into the line of sight by interstellar dust. This 
    was done by correlating the UVB against the IRAS 100 $\mu$m emission 
    (Waller {\it et al.}), or against the HI column density (Sasseen \etal and 
    Murthy {\it et al.}). Both the UIT and FAUST data, at 250 and 165 nm 
    respectively, indicate a residual UVB of about 40-100 count units 
    (cu=photons sec$^{-1}$ cm$^2$ \AA$^{-1}$ ster$^{-1}$) for negligible 
    100 $\mu$m emission or HI column density; this value is normally 
    taken as an upper limit to the extragalactic UVB (eUVB).  
 
    The levels of the extragalactic UVB are very low; according to 
    Martin (these proceedings) it is possible to account for at least 25\% 
    of the eUVB at 200 nm by the combined emission of unresolved galaxies in 
    the nearby Universe, from rest velocity to z$\approx$0.7. This is consistent 
    with the evaluation by Armand \etal (1994) of 40--130 cu, from an 
    extrapolation of the FOCA galaxy counts. The value of the eUVB at 
    $\sim$100 nm, resulting from the reconsideration of  
    long Voyager UVS observations by Murthy \etal (1996), is even more 
    intriguing. It implies an 
    eUVB consistent with zero, thus a negligible contribution by nearby 
    galaxies (to z$\approx$0.2) of photons longward of the Lyman break, and 
    also indicates that there is no significant leakage of Lyman continuum photons from 
    galaxies. This has been calculated also by Deharveng \etal (1997, 
    preprint), from a consideration of H$\alpha$ emission of galaxies.  
 
    The low sky background in the UV implies that very deep observations 
    can be made there with only modestly sized optics. This is also because  
    UV detectors are (almost) noise-free; there is no thermal noise 
    and the main sources of detector background originate from induced 
    cosmic ray events, fluorescense of optics after South Atlantic Anomaly 
    passage (in low Earth orbit), and detector hot spots. With reasonable 
    care it is possible to achieve internal noise levels of order 
    1 count sec$^{-1}$ cm $^{-2}$ of cathode and with fairly high
     quantum efficiencies (20-40\% for semi-transparent 
    cathodes and up to 60\% for opaque ones). The  internal noise level 
    is equivalent to 6.3 c.u. 
    for TAUVEX (see below) with the SF-1 filter, and is about 6$\times$ lower
    than  recent eUVB values.  
 
    The arguments presented above indicate that faint stars, and even more 
    interestingly, low surface brightness (LSB) extended objects, can be 
    detected advantageously in the UV by space experiments. This was already 
    pointed out by O'Connell (1987); the importance of an unbiased 
    survey of LSB galaxies cannot be overestimated. It is possible that a 
    large fraction of the baryon content of the Universe resides in such 
    objects (Impey \& Bothun 1997). Among the point-like objects, the more 
    interesting
    are the mixed-type binary systems where one component is a hot, evolved star.
    These systems can sometimes be detected optically only through spectroscopy, whereas a
    color-color diagram including UV would immediately show them as peculiarly
    UV-bright.
 
    \section*{UV astronomy-the reality}  

    The present knowledge of the UV sky results mostly from (a) the all-sky 
    survey by TD-1 in the early 1970s, (b)  deeper surveys of small 
    fractions of the sky by balloon, rocket, Shuttle and MIR based 
    telescopes, and (c) detailed investigation of individual 
    objects by IUE, HST, and other spectroscopic instruments.  
 
    The only all-sky survey in the UV ever performed was by the TD-1 mission 
    (Boksenberg \etal 1973). This photometric survey measured stars in four 
    spectral bands, three from 130 to 255 nm and a fourth centered at 275 
    nm. The TD-1 survey was published as a catalog with 31,215 sources; an 
    unpublished version with 58,012 sources was later produced by Landsman 
    (1984). The UV objects are mostly brighter than m$_{UV}$=8.5; this is 
    also probably the limit beyond which the UV measurements are not linear. 
    The total number of sources measured by TD-1 is similar to that of the 
    HD optical catalog !  Therefore,  the status of the knowledge of the 
    UV sky, close to the end of the second millenium,
    is like that of optical astronomy $\sim$100 
    years ago.  
 
    Among the many UV imaging missions it is worth 
    mentioning a few with particularly high sky coverage. These are the 
    S201 Moon-based experiment from NRL (Page \etal 1982), the FAUST 
    experiment (Deharveng \etal 1979; Bowyer \etal 
    1993), the SCAP-2000 and FOCA telescopes operated by the LAS Marseille group 
    (Laget 1980; Milliard \etal 1991), and the UIT missions (Stecher \etal 
    1992). The first two experiments have had very wide fields of view; 
    20$^{\circ}$ for S201 and 8$^{\circ}$ for FAUST. The latter, in 
    particular, was equipped with a modern electronic-readout detector, 
    while all the rest recorded their results on film. The FOCA project 
    operates a 40 cm telescope from a high-altitude 
    balloon; at 40+ km altitude it is possible to observe through a 
    $\sim$15 nm wide atmospheric window centered near 200 nm. The highest 
    resolution images (3") for a reasonably wide (40') FOV were obtained by 
    UIT. The data form the benchmark UV images on the morphology of nearby 
    galaxies (see papers by O'Connell, Waller {\it et al.,} Marcum, and 
    Ohl in these proceedings).  
 
    It is worth considering the capabilities of the HST in the area of UV 
    imaging; this can be done now with the WFPC2, covering a field of some 5 
    square minutes of arc, but there are significant problems with the camera 
    response. In the far-UV region the sensitivity is fairly low, while at 
    intemediate wavelengths the CCDs
    introduce significant red leaks. These problems are 
    compensated by the exquisite image quality, with the possibility to 
    resolve individual stars in nearby galaxies. The FOC is more appropriate
    for UV, but it has a much smaller field of view. To conclude,
     the HST is {\bf 
    not} a survey instrument, and was not designed specifically with high 
    performance requirements in the UV.  

\begin{table}[htb]
\begin{center}
\caption{ UV  imaging survey missions}
    \begin{tabular}{cccrrr} 
    Mission & Year & $\Omega$ (ster)& m$_L$ &  $\lambda\lambda$ (nm) & N$_{sources}$  \\ \hline 
    TD-1    & 1968-73 & 4$\pi$ &  8.8 &   150-280 & 31,215  \\ 
    S201     & 1972 & 0.96  & 11  &   125-160 & 6,266 \\ 
    WF-UVCAM & 1983 & 1.02  & 9.3 &   193     & ?    \\ 
    SCAP-2000 & 1985 & 1.88 & 13.5 &   200 & 241  \\ 
    GUV       & 1987 & 5 10$^{-3}$ & 14.5 &  156 & 52     \\ 
    GSFC CAM & 1987+ & 0.03 & 16.3 &  242 & $\sim$200  \\ 
    FOCA & 1990+ & 0.02 & 19 &  200 & $\sim$4,000   \\ 
    UIT-1 & 1990 & 3.8 10$^{-4}$ & 17 &   $\sim$270 & 2,244   \\ 
    GLAZAR & 1990 & 4.4 10$^{-3}$ & 8.7   & 164 & 489   \\
    FUVCAM & 1991 & 0.09 & 10 &   133, 178 & 1,252  \\ 
    FAUST & 1992 & 0.33 & 13.5 & 165 & 4,698 \\ 
    UIT 1+2 & 1990, 95 & 1.3 10$^{-3}$ & 19 &  152-270 & 6,000 ? \\  
    HST WFPC & 1990+ & 3.9 10$^{-4}$ & 21 &  120-300 & 50,000 ?   \\  
    MSX UVISI & 1997+ & 4$\pi$ ? & 20.0 &   180-300 & ?  \\ 
    GIMI      & 1997+ & 4$\pi$   & 13.6 &   155 & 2.5 10$^5$ \\ 
    TAUVEX & 1998+ & 0.06 & 19 &   135-270 & 10$^6$  \\ 
    XMM/OM & 1999+ & 0.05 & 20 & 150-550 & 10$^6$ ? \\

    \end{tabular}
\end{center}
 
 \end{table}

    The different instruments which performed all-sky, or partial, surveys 
    yielded a definite picture of the sky, which can be summarized as 
    follows:  
 
    \begin{enumerate}  
 
    \item There are two different stellar components to the UV sky: hot 
    massive young stars and old evolved hot stars. The two populations are 
    not co-local; the hot evolved stars reside also in the halo and in globular 
    clusters, while the young population resides in the disk.  
 
    \item A significant fraction of the diffuse UV emission is scattered 
    starlight off dust grains, even at high galactic latitudes. The dust 
    scattered UV correlates with HI and FIR emission.  
 
    \item There is a small, if any, contribution from extragalactic objects. 
    Projecting from the few galaxies of the UV-selected FOCA sample, and 
    combinig this with a likely redshift distribution, it may be possible to 
    account for a quarter or even more of the UVB by unresolved galaxies.  
 
    \item The expectation that the UV is the region with the lowest sky 
    background has been verified. In attempting to detect extended 
    objects of very low surface brightness, one must be wary of extended 
    dust patches which reflect UV 
    light from the Galaxy and may be mistaken as LSB galaxies. 
 
    \end{enumerate}  
  Table 1 summarizes information on past, present, and future UV survey
missions. It gives the solid angle covered by the survey ($\Omega$) and
the limiting magnitude achieved in the UV (m$_L$). Simple-minded
assumptions about future missions allow one to estimate how many
sources are likely to be detected by each mission (N$_{sources}$).

    In many UV survey missions it is possible to discern a 
    similar trend; they resulted only in a catalog of sources and had  
    limited scientific follow-up. This is very different from what has 
    happened in the field of high energy astrophysics (HEA), which started 
    in a very similar way. In both high energy and UV astrophysics a 
    significant driver was military technology, detectors and detection 
    alike. However, while the HEA field managed to initiate new missions one 
    after the other, with a significant number of all-sky surveys, the 
    UV was left behind. In the next section I attempt to understand why this 
    happened.  
 
    \section*{Problems of UV astronomy}  
    Many imaging experiments, in which large fractions of the sky were 
    surveyed, suffered from relatively low angular resolution and low 
    sensitivity. For instance, FAUST and FUVCAM (Carruthers \etal 1992) 
    produced images with a 
    resolution of $\sim$3'.   The identification with 
    optical counterparts in sparse regions of the sky is relatively easy, but 
    some of the identifications close to the Galactic plane, such as
    those by Schmidt \& 
    Carruthers (1993 {\it et seq.}) are problematic. These 
    are obtained by unconstrained correlations against 
    existing star catalogs, such as the SAO and the HD, and in many cases a UV 
    source is identified with a late-type star while an early type is 
    within the error ellipse but its visual magnitude is below the catalog 
    threshold. 

    Another problem with low Earth orbit missions is contamination by
    atmospheric scattered Sunlight. In case of Shuttle-based telescopes,
    an additional difficulty are the attitude jets; whenever
    these fire, wide field imagers collect many stray photons.
    Electronic readout detectors can eliminate these
    instances from the data stream when creating the final image, but this is
    not possible with film recording systems.

Finally, a major problem is conceptual; it is easy to believe that everything
to be observed by UV imaging surveys can be predicted just from enough information
gathered in the optical domain. If this were the case, ``cheap'' information
from ground-based instruments could be used to economize on ``expensive''
space adventures. In principle, this works for stars, provided one knows the
spectral type and luminosity class of an object, but is much harder for
galaxies. The situation becomes very complicated when one wants to account
properly for the very patchy galactic extinction. These difficulties
caused at least two major decisions against further advance in the UV research.

    The report of the Astronomy and Astrophysics Survey Committee, published 
    in March 1991, was one of the two 
    important decision-making events which shaped the field of observational 
    astronomy for more than one decade. Bahcall, who chaired the committee, 
    writes of the mode the decisions were taken, mostly by consensus among
    its members and within the community
    (Bahcall 1991). The survey committee assigned highest priorities for IR 
    space facilities, for large ground-based telescopes, and for the 
    millimeter array. Lower priorities were assigned to ``moderate 
    programs'', but note that the only UV mission to be included within this 
    portion of the recommendations was FUSE. Finally, the Bahcall committee 
    did not prioritize small programs, but gave three illustrative such 
    undertakings; all three are infrared studies.  
 
    Half a decade later, the report of the ``HST \& Beyond'' committee 
    (Dressler 1996) charted the  UV-O-IR astronomy well into the next 
    century. The Dressler committee identified two major goals on 
    which special effort should be expended:   detailed study of the 
    formation and evolution of normal galaxies, and  detection of 
    Earth-like planets and search for life on them. To achieve these goals 
    the committee recommended the continued operation of HST beyond its 
    designed termination in 2005, the development of the Next Generation 
    Space Telescope optimized for near-IR imaging and spectroscopy, and 
    development of space interferometry. Note, in particular, that the HST 
    would become the main UV instrument beyond 2005.
    To conclude, the only UV undertakings likely to be supported in the 
    coming decades, in a limited way, by academic or NASA 
    establishments, are FUSE and HST. It is worth considering just 
    what these missions will contribute to the general UV knowledge.  
 
    FUSE is designed for high spectral resolution (R=30,000) observations in 
    the 90.5-119.5 nm band. It will be equipped with four telescopes with 
    $\sim$36 cm apertures feeding large, dense holographic gratings, and 
    using delay-line detectors. Each telescope is optimized for a different 
    region in the full band and the effective area peaks at 105-110 nm to 
    $\sim$100 cm$^2$. This allows observation with the full spectral 
    resolution, with S/N=20, for objects as faint as 5 10$^{-14}$ erg 
    cm$^{-2}$ sec$^{-1}$ \AA$^{-1}$ ($m_{UV}\approx$12). 
    However, reaching this good S/N requires 100,000 sec 
    observations.  
 
    FUSE, being in a low Earth orbit, will not be able to access an object 
    continuously (except in the CVZ) and it is reasonable to assume a 35\% 
    duty cycle, as for HST. The mission duration is three years, thus it will 
    be possible to observe $\sim$300 faint targets (but many more brighter 
    ones).  It is clear that the basic 
    observations by FUSE will allow first rate abundance studies and 
    characterization of physical conditions in astrophysical plasmas, in the 
    Milky Way and in some nearby galaxies, but this will not be a survey
    mission.  
 
    When considering FUSE and its goals, it is worth comparing it with HUT and
    ORFEUS. HUT is a 0.9-m telescope which did low resolution spectroscopy
    (R$\approx$300) below Ly-$\alpha$ from the Shuttle bay (Davidsen et al 1992). 
    ORFEUS is a 1-m telescope which flew twice on the ASTRO-SPAS 
    platform (Kraemer \etal 1988). ORFEUS offers low and intermediate-resolution spectroscopy in 
    a spectral band similar to that of FUSE, and through the parallel IMAPS 
    spectrograph, an even higher spectral resolution, up to R=120,000. 
    IMAPS has an effective collection area of $\sim$4.5 cm$^2$ and is 
    mechanically collimated to $\sim2^{\circ}$. It is designed 
    specifically to measure the spectral line profiles of the hot ISM.  
 
    The other UV instrument for the next decades is HST. In principle, 
    its UV imaging capabiltities have always been secondary to those in the 
    visual range; only the STIS and the FOC have some pure UV 
    imaging capability, but both have tiny fields of view. 
    The Advanced Camera for Surveys will be installed during the 1999 HST 
    refurbishing mission and it is possible that then WFPC2 will be turned off. 
    The ACS is equipped with two UV channels, one with CCDs and 
    another with a STIS-like MAMA with a solar-blind CsI cathode. Both channels will 
    cover $\sim$30". We have seen here a proposal for 
    an advanced instrument for HST beyond 2002; the COS by Heap \etal This 
    would provide R=20,000 spectra in the 120-170 nm range, equivalent to 
    the high dispersion mode of the SW camera of IUE, but with a much larger 
    telescope. Note that there is no proposal for a wide-field UV imaging 
    facility with HST, although in principle something could be designed to 
    have the FOV of the WFPC2 or maybe even larger, taking advantage of the 
    possibility that WFPC2 will not be used after 1999, and replacing it on HST. 
 
    It is clear that the recommendations of the Bahcall and Dressler 
    committees effectively blocked any new initiative in the UV domain, 
    witness the unsuccessful competition for SMEX/MIDEX missions by JUNO, 
    MUSIC, HUBE, etc. This unfortunate development reflected badly in 
    prioritizing space astronomy in other countries as well; to witness, no 
    UV mission is being developed in Europe or Japan, where most space 
    activities outside of the US take place. Given this situation, it is worth 
    mentioning two UV missions being operated or prepared in the 
    US by non-academic organizations and two which piggy-back on international
    high-energy astrophysics missions.  
 
    \section*{MSX, ARGOS, SRG, and XMM}  
    {\bf MSX} is the Mid-Course Space Experiment, designed to detect and 
    characterize celestial backgrounds and atmospheric properties. It is 
    equipped with two UV imagers ({\bf UVISI}; Heffernan \etal 1996; Murthy, private
    communication), one 
    with low ($\sim$3') and another with higher ($\sim$20") resolution. The latter, in 
    particular, is capable of 
    observing in the 180-300 nm band and is probably sensitive to sources 
    brighter than m$_{UV}\approx$20 with its widest spectral 
    band. It is not clear how much of the sky will be covered by UVISI, and 
    not even when will the calibrated data become available. One source of 
    optimism is that the cryogenic material for the IR instrument was 
    exhausted ahead of time, leaving much observing time to be allocated for 
    UV studies. This is, however, only a possibility which depends on the actual 
    priorities of BMDO, who operates MSX.  
 
    {\bf ARGOS} is a USAF mission dedicated to celestial backgrounds 
    characterization, which includes a UV survey instrument from NRL. This 
    is {\bf GIMI} (Global Imaging Monitor of the Ionosphere; Carruthers \& Seeley
    1996), which consists of 
    two Schmidt cameras with 
    10$^{\circ}$.5 fields of view, imaging the spectral range form $\sim$70 
    to 200 nm in three separate bands. The calculated sensitivity of GIMI is 
    such that it will detect 13.6 mag objects, which will be imaged with 
    3'.9 resolution. At the time of this meeting, GIMI has been mated with
    the carrier platform (the ARGOS satellite) and will probably be launched
    late in 1997 or in early 1998. The important aspect of  GIMI is that
    an all-sky survey {\bf is} planned and will probably be accomplished
    within the first year in space. Another advantage is that simultaneous
    observations shall be obtained in two bands for most of the objects, and in
    three bands with short time differences. The disadvantage is that the
    angular resolution is coarse and the imaging sensitivity is not very high; 
    hopefully this will relieve the possibility of confusion with optical
    counterparts at least at high galactic latitudes.


The {\bf TAUVEX} experiment was built in Israel to provide  
deep UV imaging with reasonable resolution and wide field. It 
consists of three co-aligned 20 cm diameter telescopes imaging a 0$^{\circ}$.9
field of view with 10" resolution onto position-sensitive photon-counting
detectors. The spectral region from 130 to 290 nm is analysed with six different
bands, from one $\sim$80 nm wide to one which is only 15 nm wide.
TAUVEX is one of the instruments which will operate on board the international
observatory Spectrum X-$\gamma$ ({\bf SRG}), which is dedicated to high energy
astrophysics.

The calculated performance of TAUVEX is such that in typical SRG observations
it will detect $m_{UV}\leq$19-20 mag objects with S/N$>$5. 
The specific advantage of TAUVEX is the possibility of simultaneous
imaging in three UV bands; it is even possible, for selected objets in the
field of view, to perform fast photometry with integration times down to
10 msec. Note also that because of the special detectors, the TAUVEX
bandpasses are ``solar-blind''. In this it has much in common with the {\bf XMM
UV/Optical Monitor} (Sasseen, these proceedings). The differences are that the
XMM OM is a single 30 cm telescope, has about twice the collecting area of a 
single TAUVEX telescope, and in principle offers higher angular resolution,
up to one arcsec. In principle, the XMM OM can register very faint images,
down to m$\approx$24, but this is in the ``open'' configuration, {\it i.e.,}
UV and optical photons mixed together.

\section*{A prescription for the future}

It is clear that not much is likely to happen in the coming decade(s) unless
the UV community
manages to convince the Space Agencies that it is worthwhile to initiate  a UV
all-sky survey. A decision to proceed to a full, sensitive, all-sky survey
must be driven by the desire to discover new phenomena and to clarify puzzling
problems. I argued above that the questions of LSB galaxies and a survey for
mixed-type binaries are attractive and  solvable by a UV survey. However, 
before proceeding with planning 
such an undertaking, it would be advisable to initiate a similar one
without recourse to spacecraft technologies (and requirements); this would be much less
expensive and could be used to highlight the necessity of a space-based follow-up.

I propose that the heritage of SCAP-2000 and FOCA be used to conduct a full sky
survey in the 200 nm band using long-duration balloon flights. The history of
the FOCA experiment shows that suitable wide field and high resolution
imaging can be obtained from a high altitude balloon in this specific atmospheric
window. Long-duration balloon flights, of many days and even weeks, have been conducted
in the context of high energy and IR astrophysics.
It is likely that such modes of observation will become more popular in the
near future. Also, NASA developed a TDRSS data relay system just for
these long-duration balloon flights.

In order to assemble a long-duration mission for surveying a large fraction of the sky
in the UV one has to combine a FOCA-like telescope with a digital-readout, 
position-sensitive detector and electronics, and ensure TDRSS availability.
Observations would be conducted only at night, and during daytime the payload
will charge its batteries with satellite-like solar panels. If flights as long
as a number of weeks would be possible, as  demonstrated for Antarctic
and trans-oceanic balloons, with a few flights it will be possible to map the
entire sky at least to m$_{UV}\approx$18. This will be a large step forward, compared
with the present situation, and one which is doable for a few M\$.

Other worthwhile ventures, which can help in acquiring unique UV information, require
securing launch and operation facilities on Planetary  Astronomy missions. Such
spacecraft normally carry small-sized optics for optical-UV imaging which are
dormant during most of the cruise-phase of a mission. Despite the smallness of the 
optics, with good planning it may be possible to survey significant segments of the
sky in spectral bands which cannot be observed from balloon-borne telescopes and provide
multi-spectral UV information for m$_{UV}\leq$15 objects. Only after such an exploratory
phase it would make sense to initiate a new sky survey. In case there would be an
official decision to stop operating WFPC-2 while keeping HST operational beyond 2002, 
it is advisable to study a WFPC replacement dedicated exclusively to UV
imaging. This would offer excellent imaging with a reasonable FOV and would be
unique for studying the structure and evolution of galaxies.
      
   \section*{Acknowledgements}
I acknowledge a grant from Hughes STX  which allowed me to attend this conference.
UV studies at Tel Aviv University are supported by grants from the Ministry of
Science through the Israel Space Agency, from the Israel Academy of Sciences,
from the US-Israel Binational Science Foundation, and from the Austrian Friends of 
Tel Aviv University. I am grateful to Jean-Michel Deharveng, Alan Dressler, Jayant
Murthy, and 
Bill Waller for remarks on an early version of this paper.

    \end{document}